\documentclass[runningheads]{llncs}
\usepackage{amsmath,amssymb,amsfonts}
\usepackage{algorithmic}
\usepackage{graphicx}
\usepackage{textcomp}
\usepackage{subcaption}
\usepackage{xcolor}

\newcommand{\nop}[1]{}

\begin{document}

\title{Impact of Packetization on Network Calculus Analysis}
\author{yuming.jiang }
\date{August 2025}

\author{Yuming Jiang}
\authorrunning{Y. Jiang}
\institute{Norwegian University of Science and Technology, Trondheim, Norway 
\email{yuming.jiang@ntnu.no}}

\maketitle

\begin{abstract}
For packet-switched networks, when the packetization effect is overlooked, network calculus analysis can produce faulty results. To exemplify, network calculus analysis is applied in this paper to two basic systems that are fundamental or default settings in Time-Sensitive Networking (TSN) and Deterministic Networking (DetNet). Through counterexamples, it is revealed that for the two fundamental settings, some widely adopted, network calculus-based service characterization results, known as service curves, which ignore packetization, are faulty. In addition, for performance bounds derived from the faulty service curves, it is shown that the validity of the bounds can be arguable. In particular, the output bound, backlog bound and concatenation service curve results are shown to be also faulty: counterexamples can be constructed. By factoring the packetization effect directly into the service models, corrected service curves and performance bounds are derived for the two basic systems. These results remind that special care is needed when applying network calculus analysis to packet-switched networks. 

\end{abstract}

\section{Introduction}\label{sec-1}

Network calculus (NC) is a theory for performance guarantee analysis of communication networks \cite{Chang00}\cite{NetCal}\cite{SNC}\cite{DNC}. A key idea of network calculus is to model traffic and service processes using some bounding functions and base the analysis on them. Among the various network calculus models, service curve models play a central role, based on which various performance bounds can be derived \cite{Chang00}\cite{NetCal}\cite{SNC}\cite{DNC}. 
Since its introduction in the early 1990s \cite{Cruz91a} \cite{Cruz91b}, the network calculus has been extended and applied to various types of communication networks that are packet-switched. In particular, for IEEE 802.1 Time-Sensitive Networking (TSN) networks \cite{8021Q} and for IETF Deterministic Networking (DetNet) networks \cite{DetNet}, the theory has been extensively utilized to construct service models and compute performance bounds. Representative results include \cite{MB14}, \cite{Zhao18}, \cite{Ehsan18} and \cite{Zhao21}. A comprehensive review can be found in \cite{Zhao22}, and more recent results include \cite{Zhao24} \cite{Miserez24}. All these results are based on the service curve models of the transmission selection schemes studied, among which the constant bit rate link and strict priority scheduling are the most fundamental or default settings \cite{8021Q} \cite{DetNet}.

However, the investigation in this paper reveals that the widely adopted service curve characterization results, e.g. \cite{MB14}, \cite{Zhao18}, \cite{Ehsan18}, \cite{Zhao21}, \cite{Zhao24}, and \cite{Miserez24}, for the two fundamental settings are faulty. In particular, counterexamples can be constructed, which disapproves the results of the adopted service curves. A closer look shows that the packetization effect is overlooked in them.
The investigation is extended to examine the validity of the performance bounds derived from the faulty service curves. Counterexamples can also be constructed for output and backlog bounds as well as for the characterisation of the service curve of a concatenation system. This indicates that there is a need to update the service curve models and consequently also the performance bounds. To meet this need, corrected service curves and performance bounds for the two fundamental settings are derived, based on the idea of factoring the packetization effect directly into the service model. These results remind the  necessity of taking the packetization effect into account when applying network calculus analysis. 

The rest is organized as follows. In the next section, the system model and network calculus basics are introduced. In Section \ref{sec-3}, it is shown with counterexamples that the widely adopted service curves for the two fundamental settings are faulty due to ignoring the packetization effect. Then, a study on the impact of using such faulty service curves on performance bounds is conducted. At the end of Section \ref{sec-3}, discussion on related results from the network calculus theory is provided, which reveals that adjustments suggested by the theory for packetization have been overlooked by the faulty service curve results. In Section \ref{sec-4}, corrected service curves and performance bounds are proved. Finally, conclusions are drawn in Section \ref{sec-5}. 

\section{System Model and Network Calculus Basics} \label{sec-2}
\subsection{System Model} \label{sec-2.1}

We consider systems in a packet-switched network. In particular, we focus on data traffic transmission over a constant bit rate link and through a priority queue that competes transmission over such a link with other priority queues. 
By convention, in such a system, {\em a packet is said to have arrived (respectively, been served) when and only when its last bit has arrived (respectively, departed)} \cite{8021Q}. When a packet arrives, the packet may be queued and the buffer size for the queue is assumed to be large enough, ensuring no packet loss. The queue is FIFO and initially empty.

\begin{figure}[th!]
\centering
  \includegraphics[width=0.6\linewidth]{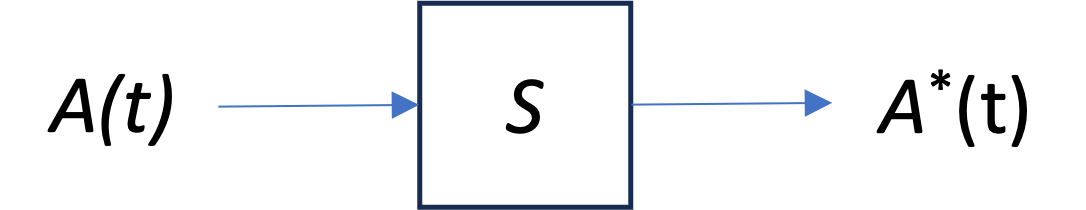} 
  \caption{System model} 
  \label{fig-sys}
\end{figure}

The system is modeled by an input traffic process $A(t)$, a service process $S$, and an output traffic process $A^{*}(t)$ as illustrated in Figure \ref{fig-sys}. Specifically, $A(t)$ denotes the cumulative amount of traffic from the input flow entering the system and $A^{*}(t)$ the cumulative amount of traffic from the flow leaving the system, up to time $t$ (excluded). By convention, we adopt $A(0) = A^{*}(0) = 0$. In addition, we define $A(s,t) \equiv A(t)-A(s)$ and $A^{*}(s,t) \equiv A^{*}(t)-A^{*}(s)$, which respectively denote the amount of input traffic and the amount of output traffic in period $[s, t)$. Since traffic at $t$ is excluded, $A(t,t)=0$ by convention and so is $A^{*}(t,t)=0$.

In addition, we can also model the input and output processes of the flow using marked-point processes, where packetization of traffic is explicitly taken into account \cite{Chang00}. Specifically, for the input, the marked point process $(\overrightarrow{a}, \overrightarrow{l})$ consists of two sequences of variables $\overrightarrow{a}=\{a(n), n=0, 1, 2, ...\}$ and $\overrightarrow{l}=\{l(n), n=0, 1, 2, ...\}$, where $a(n)$ denotes the arrival time of packet $n$ and $l(n)$ its length. For the output, a similar marked point process is defined which is $(\overrightarrow{d}, \overrightarrow{l})$  with $\overrightarrow{d} =\{d(n), n=0, 1, 2, ...\}$, where $d(n)$ denotes the departure time of the $n$-th packet from the system. Here, the packet $0$ is a virtual packet, and by convention $a(0)=0$, $l(0)=0$ and $d(0)=0$.  

The backlog at time $t$, denoted as $B(t)$, is: 
\begin{equation}
B(t) = A(t) - A^*(t).
\end{equation}
The delay of packet $n (\ge 1)$, denoted by $D(n)$, is:
\begin{equation}
D(n) = d(n) -a(n).
\end{equation}
In network calculus, a related delay concept, called virtual delay, has been adopted. Specifically, the virtual delay at time $t (>0)$, denoted as $D(t)$, is defined as \cite{Chang00}\cite{NetCal}\cite{SNC}\cite{DNC}:  
\begin{equation}
D(t) = \inf\{\tau \ge0: A(t) \le A^{*}(t+\tau)\}. 
\end{equation}
 
\subsection{Network Calculus Basics} \label{sec-2.2}

Let $\mathbf{\mathcal{F}}$ denote the set of nonnegative nondecreasing functions, and $\mathbf{\mathcal{F}}_0$ its subset with $f(0) = 0$. By their definitions, $A(\cdot)$ and $A^{*}(\cdot)$  are both in $\mathbf{\mathcal{F}}_0$. 

\begin{definition}\label{def-ac}
A flow $A$ is said to have an arrival curve $\alpha \in \mathbf{\mathcal{F}}$, if for all $0\le s \le t$ , the traffic $A(s,t)$ is upper-constrained by \cite{NetCal}:
\begin{equation}
A(s, t) \le \alpha(t-s).
\end{equation}
\end{definition} 

\begin{definition}\label{def-sc}
A system is said to provide a service curve $\beta \in \mathbf{\mathcal{F}}_0$, if for any time $t\ge 0$, there exists some time $s \in [0, t]$ such that 
$A^{*}(t) \ge A(s) + \beta(t-s)$ 
or equivalently, there holds for all $t \ge 0$ \cite{NetCal}, $$A^{*}(t) \ge A \otimes \beta(t) \equiv \inf_{0 \le s \le t}\{A(s) + \beta(t-s)\}.$$
\end{definition}

\begin{theorem}\label{th-1}
If the input flow has an arrival curve $\alpha$ and the system provides to the input flow a service curve $\beta$, there hold \cite{NetCal}:\\
(i) the output has an arrival curve $\alpha^{*}(t)\equiv \sup_{u \ge 0}\{\alpha(u+t) - \beta(u)\} $, i.e. $\forall s, t \ge 0$,  
\begin{equation}\label{min-out} 
A^{*}(s, s+t) \le  \alpha^{*}(t);
\end{equation}
(ii) the backlog $B(t)$ is upper-bounded by, $\forall t \ge 0$,  \\
\begin{equation}\label{min-bb} 
B(t) \le \sup_{t \ge 0}\{\alpha(t) - \beta(t)\}; 
\end{equation}
(iii) the virtual delay $D(t)$ is upper-bounded by, $\forall t \ge 0$: 
\begin{equation}\label{min-db} 
D(t) \le \sup_{t \ge 0} \inf\{\tau \ge 0: \alpha(t) \le \beta(t+\tau) \}. 
\end{equation} 
\end{theorem}

As an example, suppose that the input flow is constrained by a token bucket with token generating rate $\rho$ and bucket size $\sigma$ \cite{Cruz91a} or has an arrival curve $\alpha(t) = \rho t +\sigma$. In addition, the system provides to the input a latency-rate service curve $\beta(t)=R(t-T)^{+}$, where $R$ is called the rate term and $T$ the latency term. If $\rho \le R$, the bounds from Theorem \ref{th-1} can be written as
\begin{eqnarray}
\alpha^{*}(t) &=& \rho t + (\sigma + \rho T)\\
    B(t) &\le& \sigma + \rho T\\
    D(t) &\le& \frac{\sigma}{R}+T
\end{eqnarray}

In network calculus analysis, an important technique for finding the service curve characterization of a system makes use of the concept of strict service curve and its relation with the service curve model as summarized below. 

\begin{definition}\label{def-ssc}
A system is said to provide a strict service curve $\beta \in \mathbf{\mathcal{F}}_0$, if during any backlogged period of length $t$, the output satisfies \cite{NetCal} 
$A^{*}(t) \ge \beta(t)$. 
\end{definition}

\begin{proposition}\label{pro-ssc}
    If $\beta (\in \mathbf{\mathcal{F}}_0$) is a strict service curve of a system, it is also a service curve of the system \cite{NetCal}.  
\end{proposition}

For a concatenation system, the following result is important for finding its service curve characterization. 

\begin{theorem}\label{th-2}
    Consider a flow that traverses two systems $S_1$ and $S_2$ in sequence. If each system $S_i, (i=1,2)$ offers a service curve of $\beta_i$ to its input, then the concatenated system offers a service curve of $\beta_1 \otimes \beta_2$ to the flow \cite{NetCal}.
\end{theorem}

\section{Impact of Packetization on Network Calculus Analysis} \label{sec-3} 

In this paper, we focus on two fundamental transmission selection settings in a packet-switched network. One is a (work-conserving) constant bit rate link, and the other is a strict priority queue for packet transmission on such a link. The former is the foundation of any other transmission selection algorithm, while the latter is commonly implemented, e.g. as the default setting for time-sensitive networking \cite{8021Q}. 
For the two fundamental cases, the following service curve results have been widely applied, e.g. in \cite{MB14} \cite{Zhao18} \cite{Ehsan18} \cite{Zhao21} \cite{Zhao22} \cite{Zhao24} and \cite{Miserez24}: 
\begin{itemize}
 \item A (work-conserving) link with constant bit rate $c$ provides a (strict) service curve $c t$.
 \item When (non-preemptive) strict priority is applied on the link, the highest priority queue receives a (strict) service curve $c(t-\frac{l^{M_l}}{c})^{+}$, where $l^{M_l}$ denotes the maximum packet length of all lower priority queues and $(x)^{+}\equiv \max\{x, 0\}$. 
\end{itemize}

Unfortunately, as shown in the following Section \ref{sec-3.1}, both are faulty.

\subsection{Faulty Service Curves}\label{sec-3.1}

\subsubsection{The Constant Bit Rate Case: }

The intuition behind using $ct$ as a (strict) service curve of the system is as follows. Since the link is work-conserving, during any backlogged period, the link outputs traffic (in bits) at a rate $c$ and hence has a strict service curve $ct$, with which $ct$ as a service curve can also be concluded from Proposition \ref{pro-ssc}. While this sounds straightforward, the following example and discussion highlight that special care is needed when packetization effect needs to be taken into account. 

\begin{figure}[th!]
\centering
  \includegraphics[width=0.8\linewidth]{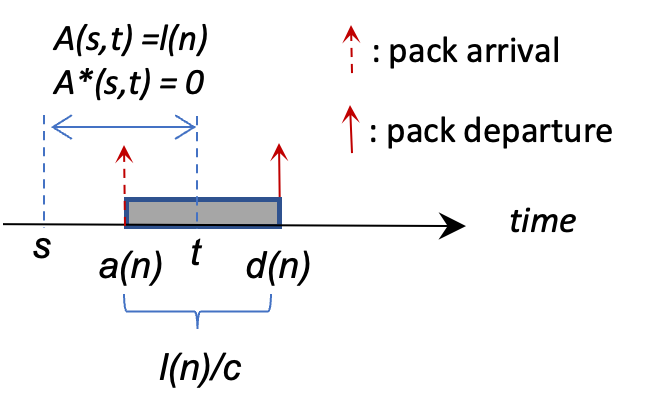} 
  \caption{Impact of packetization on the service of a constant bit rate link} 
  \label{fig-pkt}
\end{figure}

Consider the transmission of a packet $n$ on the link as illustrated in Figure \ref{fig-pkt}. Without considering the packetization effect, we would have $A(s,t)=A^*(s,t)=c(t-a(n))$ (in bits). However, under the packet model, taking packetization into account, we actually have $A(s,t)=l(n)$ while $A^*(s,t)=0$. The former is because at time $a(n)$ which is within $[s,t)$, the (last bit of the) packet has arrived so that transmission can start. In contrast, for $A^*(s,t)$, $d(n)$ is not within the period, which means that the last bit of the packet has not finished transmission, so the packet is not considered to have been transmitted or served in the period considered $[s,t)$. Because $0 \le c(t-a(n)) \le l(n)$, we have $A^*(s, t) \le c(t-s)$, which contracts the definition of strict service curve if $ct$ were used as a strict service curve. 

The discussion above disapproves the validity of using $ct$ as a strict service curve. However, the question of whether $ct$ is a valid service curve remains. For this, we construct a case where $A^*(t) \ge A \otimes \beta(t)$ does not hold for $\beta(t) = ct$. 

Consider the first packet. Let $a(1)$ be its arrival time and $l(1) (>0)$ its length. Then, for any time $t \in [a(1), a(1)+\frac{l(1)}{c})$, the period $(a(1), t)$ is backlogged because the packet has arrived but has not left. In addition, we have: 
\begin{eqnarray}
 A \otimes \beta(t) &=& \inf_{0 \le s \le t}\{A(s) + c(t-s)\} \nonumber\\
    &=& \min \left\{\inf_{0 \le s < a(1)}\{A(s) + c(t-s)\}, \inf_{a(1) \le s \le t}\{A(s) + c(t-s)\} \right\}  \nonumber\\
    &=& \min \{ c(t-a(1)), l(1)\}  \nonumber\\    
    &=& c(t-a(1)) >0
\end{eqnarray}
Note that $A^*(t)=0$ because a packet is considered to have been served only when its last bit has left, the first packet has not left at $t (<a(1) + \frac{l(1)}{c})$ and therefore the packet cannot be counted in $A^*(t)$. In other words, for any $t \in [a(1), a(1)+\frac{l(1)}{c})$, $A \otimes \beta(t) > A^*(t)$, which contradicts the requirement of $A^*(t) \ \ge A \otimes \beta(t), \forall t\ge 0$ for $\beta$ to be a service curve. Proposition \ref{pr-cr} is now concluded. 

\begin{proposition}\label{pr-cr}
For a system with constant bit rate $c$ (in bps), $c  t$ is neither a strict service curve nor a service curve. 
\end{proposition}

\subsubsection{The Strict Priority Case:}  
For the (non-preemptive) strict priority case, the intuition behind using $c(t- \frac{l^{M_l}}{c})^{+}$ as a (strict) service curve is as follows. In the formulation, $\frac{l^{M_l}}{c}$ factors in the non-preemption effect, which is, upon arrival, a highest priority packet will have to wait for the packet under transmission to complete even though the packet under transmission is from a lower priority queue. However, as discussed for Proposition \ref{pr-cr}, when packetization is taken into account, $ct$ is not a (strict) service curve for the link. In particular, factoring the non-preemption effect and adding $\frac{l^{M_l}}{c}$ at $a(n)$ in Figure \ref{pr-cr}, the same analysis taking packetization into account can be conducted for the highest priority queue, from which we can consequently conclude Proposition \ref{pr-sp}.

\begin{proposition}\label{pr-sp}
For the highest priority queue on a link with constant bit rate $c$, $c (t-\frac{l^{M_l}}{c})^{+} $ is neither a strict service curve nor a service curve.
\end{proposition}

\subsubsection{Implications: } \label{sec-3.2} 

The impact of packetization on the service curve characterization results as investigated above and summarized in Propositions \ref{pr-cr} and \ref{pr-sp} has two immediate implications: 

\begin{itemize}
\item The various performance bounds derived from the faulty service curves, such as output arrival curve, backlog and delay bounds, must be re-examined for their validity and possibly updated.  

\item For any complex setting that builds on the two fundamental cases, if its service curve result implies a faulty service curve for either of the two fundamental cases as indicated in Propositions \ref{pr-cr} and \ref{pr-sp}, the service curve result for the complex setting as well as the correspondingly derived performance bounds will need to be re-examined and possibly updated. 
\end{itemize}

For the latter, since a fundamental case is a special case of the complex setting, if the result for the special case is faulty, the result for the complex setting cannot hold either. For the former, it is worth highlighting that the constant bit rate case is implied in the strict priority case (SP). Specifically, it is a special case of SP with a single queue in the system. 

\subsection{Impact on Performance Bounds} \label{sec-3.2} 

In this subsection, we investigate the impact of a faulty service curve on the correspondingly obtained performance bound results. By constructing counterexamples, we show that the faulty service can lead to faulty output bound, faulty backlog bound and faulty concatenation service curve, while for delay bound the same conclusion cannot be made. We focus on the constant bit rate case. Because it is the most fundamental setting and is a special case of SP, the related conclusion on the invalidity of a bound is also applicable to the SP case. 

\subsubsection{Faulty output bound: }

\begin{figure}[th!]
        \centering
        \includegraphics[width=0.6\textwidth]{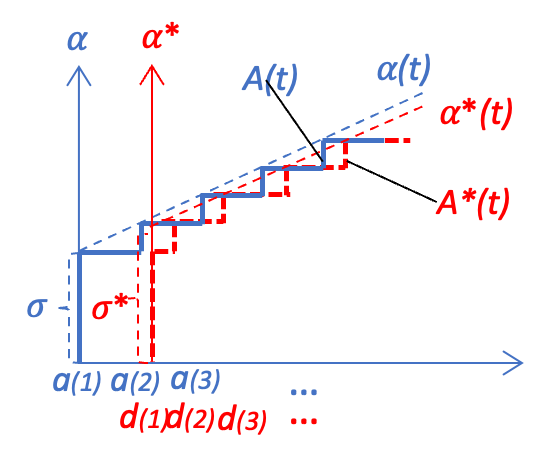}
        \caption{Input-output: Input-related in blue and output-related in red}
  \label{fig-io}
\end{figure}

To ease discussion, a simple input case is considered. Specifically, the input flow sends packets periodically to the system. Let $\tau$ denote the interval between two adjacent packets. All, except packet 1, have the same length $l$, while the length of packet 1 is $\sigma (>l)$. For stability, assume $\frac{l}{\tau} < c$. For this input case, it can be verified that it has an arrival curve $\alpha(t) = \frac{l}{\tau}t+\sigma$. An illustration of $A(t)$ and $\alpha(t)$ is presented in Fig. \ref{fig-io}, where the output $A^*(t)$ and an output arrival curve $\alpha^*(t)$ that uses the same slope as $\alpha$ are also illustrated. (Remark: For $\alpha$ and $\alpha^*$, they correspond to intervals starting from $a(1)$ and $d(1)$ respectively.)
For the output, applying $\beta(t)=ct$ as a service curve to Theorem \ref{th-1} would give $A^*(s, s+t) \le \frac{l}{\tau}t+\sigma = \alpha(t)$, $\forall s, t \ge 0$, or in other words $\alpha^{*}=\alpha$. However, as illustrated in Fig. \ref{fig-io}, $\alpha^*(t)$ needs to have a higher $\sigma^*$ than the initial $\sigma$. In other words, the output is not constrained by $\alpha(t)$ and therefore using $\alpha$ as an upper bound on the output, based on derivation from the faulty service curve, is wrong. 

\subsubsection{Faulty backlog bound: }
For backlog, applying $\beta(t)=ct$ as service curve to Theorem \ref{th-1}  gives $B(t) \le \sigma$, $\forall t \ge 0$. However, this backlog bound $\sigma$ is incorrect. Consider the same case as illustrated in Fig. \ref{fig-io}. In particular, before packet 1 finishes, packet 2 has arrived. Hence, $B(t)$ for $t = d(1)_{-}$, i.e. the time just before packet 1 has finished its transmission, the backlog in the system is $\sigma + l$ where $\sigma$ is the size of packet 1 and $l$ is that of packet 2. Since $\sigma + l > \sigma$, the backlog bound $\sigma$ from the faulty service curve is therefore incorrect.  

\subsubsection{Delay bound: }
For delay, applying $\beta(t)=ct$ as service curve to Theorem \ref{th-1} gives $D(t) \le \frac{\sigma}{c}$, $\forall t \ge 0$. Surprisingly, unlike that the output bound and backlog bound derived from the faulty service curve are also faulty, no similar conclusion can be made for the delay bound. This is because $\frac{\sigma}{c}$ is actually a valid delay bound, which has been proved but using another service-modeling technique \cite{GR}. More remarks related to this are provided in Section \ref{sec-3.3}.  

\subsubsection{Faulty concatenation service curve: } 
Assume a flow traverses two switches $S_1$ and $S_2$ in sequence where the corresponding output links have constant bit rate $c_1$ and $c_2$ respectively. Let $\beta_{i}(t)=c_i t, (i=1,2)$. If they were correct service curves for $S_1$ and $S_2$ respectively, we would have  $\beta_{1} \otimes \beta_{2} (t) = \min\{c_1, c_2\} t$ to be a service curve for the concatenation system. Then, if the input has an arrival curve of $\rho t + \sigma$ with $\rho \le \min\{c_1, c_2\}$, the delay of any packet would be upper-bounded by $\frac{\sigma}{ \min\{c_1, c_2\}}$, for which, however, counter examples can be constructed. Specifically, for simplicity, consider the case where all packets have the same length $l$, $\sigma=l$ and $c_1=c_2\equiv c$. Then we have $\frac{\sigma}{ \min\{c_1, c_2\}} = \frac{l}{ c}$. Furthermore, it is easily verified that for any packet, its delay is at least $2 \times \frac{l}{c}$ due to its transmission time of $\frac{l}{ c}$ on both systems, which is larger than $\frac{l}{c}$. This implies that concatenation of faulty service curves leads to a faulty concatenation service curve for the concatenated system. 

\subsection{Remarks}\label{sec-3.3}

It is worth highlighting that the service curves in the same forms for similar systems have been introduced in the network calculus theory but under different contexts.  
\begin{itemize}
    \item In \cite{Chang00}, discrete time is adopted and {\em the service rate $c$ is in the number of packets per unit time}. Under these assumptions, $ct$ is proved to be a service curve, called an $f$-server in \cite{Chang00},  for the constant rate server and for the highest priority queue. 
    \item In \cite{NetCal} and \cite{DNC}, the analysis and results do not depend on whether the time model is discrete or continuous, which is also the case in the present paper. In \cite{NetCal} and \cite{DNC}, by defining {\em the amount of service counted at the bit level}, $ct$ is shown to be a strict service curve for the constant bit rate case and so is $c(t-\frac{l^{M_l}}{c})^{+}$ for the highest priority queue on such a link (cf. Proposition 1.3.4 in \cite{NetCal} and Section 1.2 in \cite{DNC}). 
\end{itemize}

\begin{figure}[th!]
\centering
  \includegraphics[width=0.9\linewidth]{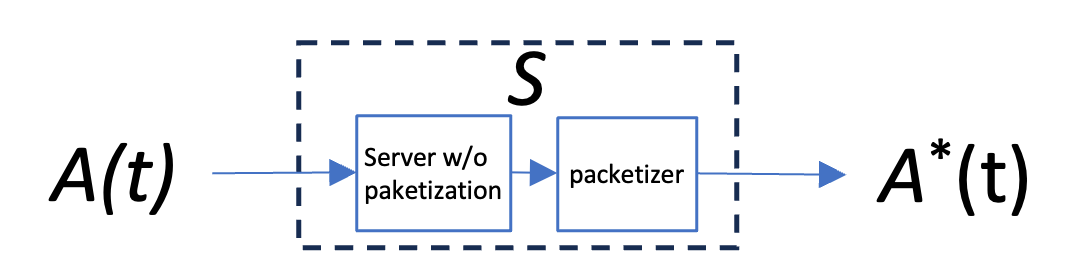} 
  \caption{System model: Packetizer} 
  \label{fig-packetizer}
\end{figure}

To account for the effect of packetization, particularly that a flow is composed of a sequence of (variable-length) packets and that the scheduling algorithm is at the packet level while not at the bit level, a novel concept called {\em packetizer} is introduced in \cite{Chang00} and also adopted in \cite{NetCal} and \cite{DNC}. Its idea is to treat the system as the concatenation of two conceptual subsystems, as shown in Fig. \ref{fig-packetizer}. One is the bit-by-bit system, which is the part of the system before packetization is taken effect, and the other is a packetizer that assembles output traffic from the previous subsystem into packets. 

With the packetizer concept, it is highlighted in \cite{Chang00}, \cite{NetCal} and \cite{DNC} that additional terms may need to be added to the service curve  and performance bound results. In particular, the service curve, the backlog bound, and the concatenation service curve must be adjusted, although the packetizer is shown to not increase the maximum packet delay \cite{Chang00}, \cite{NetCal} \cite{DNC}. Unfortunately, this seems to have been largely overlooked in the recent application of network calculus analysis, e.g., in \cite{MB14}, \cite{Zhao18}, \cite{Ehsan18}, \cite{Zhao21}  \cite{Zhao22}, \cite{Zhao24} and \cite{Miserez24}, where the faulty service curves are directly adopted without considering the packetization effect. 

Unlike relying on the concept of packetizer to account for the packetization effect in \cite{Chang00}, \cite{NetCal}, and \cite{DNC}, we take packetization into direct consideration when modeling the arrival and service processes. More specifically, as introduced in Section \ref{sec-2}, we adopt the convention that {\em a packet is said to have arrived (respectively, been served) when and only when its last bit has arrived (respectively, departed)} and hence model the arrival and departure processes as marked point processes. Based on this idea, we conduct network calculus analysis and present corrected service curves and performance bounds for the two fundamental cases in the next Section \ref{sec-4}.

\section{Corrected Service Curves and Performance Bounds}\label{sec-4}

\subsection{The Constant Bit Rate Case}
For the constant bit rate case, corrected service curves are presented in Proposition \ref{pr-cr2}. 

\begin{figure}[th!]
\centering
  \includegraphics[width=0.8\linewidth]{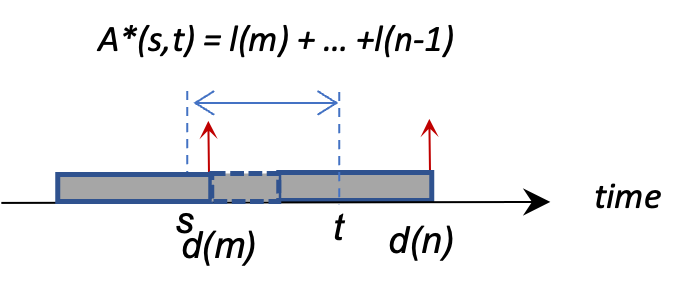} 
  \caption{Service of a link during a backlogged period} 
  \label{fig-ssc4link}
\end{figure}

\begin{proposition}\label{pr-cr2}
    A system with constant bit rate $c$ offers a strict service curve and a service curve $c (t - \frac{l^M}{c})^{+}$, where $l^M$ denotes the maximum packet length in the system. 
\end{proposition}
\begin{proof}
    Consider any backlogged period $(s, t]$ as shown in Figure \ref{fig-ssc4link}, and let $n_0$ denote the packet whose arrival starts the backlogged period. Clearly, we have $s \ge a(n_0)$. Without loss of generality, suppose that packet $m$ is the first packet that leaves after time $s$, and $n$ the first packet that departs after time $t$. By their definitions, we must have $d(m-1) < s \le d(m)$ and $d(n-1) \le t < d(n) $. Since the system is backlogged during $(s, t]$, it must also be so during (i) $(s,d(m)]$, (ii) $(d(m),d(n-1)]$ and (iii) $(d(n-1), t]$. Because the system has a constant bit rate $c$, during any backlogged period of duration $\tau$, the amount of traffic it serves / transmits is $c \tau$ (in bits). We hence have $c(d(m)-s) \le l(m)$, $c[d(n-1)-d(m)]=\sum_{k=m+1}^{n-1}l(k)$, and $c[t-d(n-1)]\le l(n) \le l^M$. Since $A^{*}(s, t)= \sum_{k=m}^{n-1}l(k)$ as illustrated by Fig. \ref{fig-ssc4link}, we have
    \begin{eqnarray}
        A^{*}(s, t) &=& l(m) + c[d(n-1)-d(m)] \nonumber \\
        &\ge& c[d(n-1)-s] \ge c(t-s - \frac{l^M}{c})
    \end{eqnarray}
which, together with the fact that $A^{*}(s, t) \ge 0$, ends the proof. 
\end{proof}

With Proposition \ref{pr-cr2}, various performance bounds can immediately be derived from Theorem \ref{th-1}. As a specific example where the input is constrained by a token-bucket arrival curve, they are summarized in Corollary \ref{cor-cr2}.

\begin{corollary}\label{cor-cr2}
    Consider a traffic flow transmitting over a link that has a constant bit rate $c$. If the traffic has a token-bucket arrival curve $\alpha(t) = \rho t +\sigma$ with $\rho \le c$, 
    there hold:\\ 
(i) the output has an arrival curve $\rho t +\sigma + \rho \frac{l^M}{c}$, i.e. $\forall s, t \ge 0$, 
\begin{equation}\label{min-bb} 
A^{*}(s, s+t) \le \rho t + \sigma + \frac{\rho}{c}l^{M}; 
\end{equation}
(ii) the backlog $B(t)$ is upper-bounded by, $\forall t \ge 0$, 
\begin{equation}\label{min-bb} 
B(t) \le \sigma + \frac{\rho}{c}l^{M}; 
\end{equation}
(iii) the virtual delay $D(t)$ is upper-bounded by, $\forall t \ge 0$: 
\begin{equation}\label{min-db} 
D(t) \le \frac{\sigma + l^M}{c}.
\end{equation} 
\end{corollary}

In addition, with Proposition \ref{pr-cr2} and Theorem \ref{th-2}, a corrected concatenation service curve can be found for the counterexample case discussed in Section \ref{sec-3.2}. 

\begin{corollary}
      Consider a flow that traverses two switches in sequence. The corresponding output links have constant bit rate $c_i, (i=1,2)$ respectively. The concatenation system offers to the flow a service curve of $\min\{c_1, c_2\}(t-\frac{l^M}{c_1}-\frac{l^M}{c_2})^{+}$. 
\end{corollary}

\subsection{The Strict Priority Case}
For the strict priority case, a corrected strict service curve, which is also a service curve, is presented in Proposition \ref{pr-sp2}. For the proof, it is similar to that of Proposition \ref{pr-cr2}. The key difference is that $s$ may be within the transmission period of a lower priority packet that has started its transmission and cannot be preempted by the arrival of the highest priority packet $n_0$. In such a scenario, we have $c(s-d(n_0)) \le l(n_0) + l^{M_l}$, $c(d(n-1)-d(n_0)) = \sum_{k=n_0+1}^{n-1}l(k)$, and $c[t-d(n-1)]\le l(n) \le l^M$. From these, we obtain 
    \begin{eqnarray}
        A^{*}(s, t) &=& \sum_{k=n_0}^{n-1}l(k) 
        \ge c[d(n-1)-s-\frac{l^{M_l}}{c}] \ge c(t-s - \frac{l^M}{c}-\frac{l^{M_l}}{c})
    \end{eqnarray}
where $l^{M}$ and $l^{M_l}$ respectively denote the maximum packet length of the considered highest priority queue and the maximum packet length of all lower priority queues. For other scenarios, where the arrival of the highest priority packet $n_0$ does not see any ongoing transmission of a lower priority packet, they are the same as in the proof of Proposition \ref{pr-cr2} and hence $A^{*}(s, t) \ge c(t-s - \frac{l^M}{c})$. Combining all scenarios and the fact that $A^{*}(s, t) \ge 0$, Proposition \ref{pr-sp2} can be concluded.  

\begin{proposition}\label{pr-sp2}
For the highest priority queue in a system with constant service rate $c$ (in bps), $c (t-\frac{l^{M}}{c} -\frac{l^{M_l}}{c})^{+} $ is both a strict service curve and a service curve.
\end{proposition}

Similarly, with Proposition \ref{pr-sp2}, various performance bounds can also be derived from Theorem \ref{th-1}. Specifically, if the input is constrained by a token-bucket arrival curve, the corresponding bounds are summarized in Corollary \ref{cor-sp2}. 

\begin{corollary}\label{cor-sp2}
    Consider a traffic flow competing with other flows to transmit on a link that has a constant bit rate $c$, and the considered flow is given the highest priority. If the traffic of the highest priority flow has a token-bucket arrival curve $\alpha(t) = \rho t +\sigma$ with $\rho \le c$, 
    there hold:\\ 
(i) the output has an arrival curve $\rho t +\sigma + \rho \frac{l^M+l^{M_l}}{c}$, i.e. $\forall s, t \ge 0$, 
\begin{equation}\label{min-bb} 
A^{*}(s, s+t) \le \rho t + \sigma + \frac{\rho}{c}(l^{M}+l^{M_l}); 
\end{equation}
(ii) the backlog $B(t)$ is upper-bounded by, $\forall t \ge 0$,  
\begin{equation}\label{min-bb} 
B(t) \le \sigma + \frac{\rho}{c}(l^{M}+l^{M_l}); 
\end{equation}
(iii) the virtual delay $D(t)$ is upper-bounded by, $\forall t \ge 0$: 
\begin{equation}\label{min-db} 
D(t) \le \frac{\sigma + l^M + l^{M_l}}{c}.
\end{equation} 
\end{corollary}

Similarly, with Proposition \ref{pr-sp2} and Theorem \ref{th-2}, a service curve can be found for the concatenation of two priority systems.   

\begin{corollary}
      Consider a flow that traverses two systems in sequence. Both systems adopt strict priority scheduling to share among flows the service rate (in bps) $c_i, (i=1,2)$, which is constant. The flow is given highest priority at both systems. Then, the concatenation system offers the flow a service curve of $\min\{c_1, c_2\}(t-\frac{l^M+l^{M_l}}{c_1}-\frac{l^M+l^{M_l}}{c_2})^{+}$. 
\end{corollary}

\subsection{Remarks}

The discussion in Section \ref{sec-3.3} has indicated that the idea of packetizer has been adopted in \cite{Chang00}\cite{NetCal}\cite{DNC} to account for the packetization effect. In particular, for the corresponding bounds as shown in Propositions \ref{pr-cr2} and \ref{pr-sp2}, the suggested adjustments can be found in \cite{Chang00}\cite{NetCal}\cite{DNC}. 

More specifically, for the constant bit case, the service curve is adjusted to be the same as shown in Proposition \ref{pr-cr2}, and the service curve for the concatenation system, the adjustment is also the same. In addition, the output is upper-bounded by, $\forall s, t \ge 0, A^{*}(s, s+t) \le \rho t + \sigma +l^M$ and the backlog is upper-bounded by $\forall t\ge0, B(t) \le \sigma +l^M$, while for virtual delay, it is upper-bounded by $\forall t\ge0, D(t) \le \frac{\sigma}{c}$. Comparing with the bounds shown in Corollary \ref{cor-cr2}, our output and backlog bounds are tighter. However, for virtual delay, the bound derived from ignoring the packetizer effect is better. For the strict priority case, similar comparison can be conducted, and it is also observed that the packetizer approach gives the same service curves, looser output and backlog bounds, but tighter delay bound. 

A final remark is that the approach of taking the packetization effect directly into account in service-modeling has been exploited in the network calculus literature, e.g. in \cite{Chang00} \cite{GR} and \cite{Jiang24}. Recent work \cite{Jiang24} particularly indicates that the same delay bounds can also be derived for the two basic systems using this approach. For a more detailed discussion, which is beyond the scope of the present paper, interested readers are referred to \cite{Chang00} \cite{GR} and \cite{Jiang24}. 

\section{Conclusion}\label{sec-5}

Network calculus analysis has been applied to two basic systems that are fundamental or default settings in time-sensitive networks. The focus has been on analyzing their service curve characterizations and performance bounds. It was shown through counterexamples that the service curve s, which have been widely adopted in some recent literature for the two fundamental settings, are actually faulty. Additionally, it was also shown that for the output bound, backlog bound and concatenation service curve results derived from the faulty service curves, counterexamples can also be constructed. This implies that both the service curve characterizations of the two fundamental settings and the correspondingly derived performance bounds need to be adjusted or re-investigated to account for the packetization effect. To this end, the service curves and performance bounds were corrected by exploring the idea of factoring packetization directly into service-modeling. The results exemplify not only that if the packetization effect is overlooked, network calculus analysis can produce faulty results, but also how the packetization effect can be factored in the analysis.

\bibliographystyle{unsrt}
\bibliography{nc-qt}
\end{document}